%% file: iclr2023_conference.tex
\documentclass{article} % For LaTeX2e
\usepackage{iclr2023_conference,times}

\input{math_commands.tex}

\usepackage[justification=centering]{caption}
\usepackage{caption}
\usepackage{hyperref}
\usepackage{url}
\usepackage{algorithm}
\usepackage{algorithmic}
\usepackage{graphicx}
\usepackage{hyperref}
\floatname{algorithm}{Procedure}

\title{UnifySpeech: A Unified Framework for Zero-shot Text-to-Speech and  Voice Conversion}

\author{Haogeng Liu\footnotemark[1], Tao Wang\footnotemark[1], Ruibo Fu\footnotemark[2], Jiangyan Yi\footnotemark[2], Zhengqi Wen\footnotemark[2] \\
Chinese Academy of Science\\
Beijing, China \\
\And
Jianhua Tao\footnote[1]{Demo for this paper can be found atxxx} \\
Department of Automation, Tsinghua University \\
School of Artificial Intelligence, University of Chinese Academy of Sciences, China \\
Beijing, China \\
}

\iclrfinalcopy % Uncomment for camera-ready version, but NOT for submission.

\begin{document}

\maketitle

\begin{abstract}
Text-to-speech (TTS) and voice conversion (VC) are two different tasks both aiming at generating high quality speaking voice according to different input modality. Due to their similarity, this paper proposes UnifySpeech, which brings TTS and VC into a unified framework for the first time. The model is based on the assumption that speech can be decoupled into three independent components: content information, speaker information, prosody information. Both TTS and VC can be regarded as mining these three parts of information from the input and completing the reconstruction of speech. For TTS, the speech content information is derived from the text, while in VC it's derived from the source speech, so all the remaining units are shared except for the speech content extraction module in the two tasks. We applied vector quantization and domain constrain to bridge the gap between the content domains of TTS and VC. Objective and subjective evaluation shows that by combining the two task, TTS obtains better speaker modeling ability while VC gets hold of impressive speech content decoupling capability. \par
\end{abstract}

\section{Introduction}
Cloning the voice of the target speaker is an attractive technology, which can be applied to various scenes~ \citep{sisman2020overview}, such as entertainment creation, personalized mobile assistants, security field, etc. The most ideal voice cloning operation is to give only one relatively short speech of the unseen target speaker as a reference and then any speech of the target speaker can be synthesized, which is called zero-shot voice clone. In the speech research community, text-to-speech (TTS) and voice conversion (VC) are two mainstream ways to realize voice clone~\citep{sisman2020overview}. Therefore, a variety of techniques for zero-shot TTS and VC have been proposed recently~\citep{gorodetskii2022zero,tang2021zero}.

However, although TTS and VC techniques are two important ways of voice clone with same output form, the research of TTS and VC is more or less independent. There isn't much interaction between them. But they are both speech synthesis task. In terms of speech generation, we categorize the information of the target speaker's speech into three kinds of information: (1) speech content, the characters of phonemes or phonetic posteriorgram (PPG) in voice conversion, represents the content of the speech. (2) speaker information, which represents the characteristics of speakers, is related to the speaker's articulation organ. (3) prosody information, which covers the intonation, stress, and rhythm of speech. According to FastSpeech2~\citep{ren2020fastspeech}, pitch, energy and duration information can reflected them certainly. As TTS extracts speech content directly from phonemes, it is easier to obtain content information irrelevant to speaker than VC. As VC encounters more speakers, it's possible to obtain more robust speaker modeling ability. So, by integrating TTS and VC into a unified framework and combining their training data, it can help the model learn these three kinds of information better.

Unfortunately, investigating TTS and VC in the same framework is challenging, as the speech content are extracted from different modality. Specifically, the speech content in TTS is obtained through phoneme information while the phonemes and speech in TTS are unequal sequences, needing attention mechanism to align them. However, the attention mechanism is often affected by the speaker's information, so it is impossible to learn the speech content representation completely irrelevant to the speaker. While in VC, the source speech and target speech are aligned in speech content, so the speech content can be extracted directly from the source speech, which is different from the TTS. In contrast to speech content information, speaker information and prosody information can be modeled using the same network in TTS and VC. Therefore, the difficulty here is to keep speech content for TTS and VC consistent. With the development of speech synthesis, the recently proposed Adaspeech2~\citep{yan2021adaspeech} can combine text information with speech information, which can effectively decouple the speech content from the input. The unified framework becomes possible.

Overall, the main contributions of this paper are:
\begin{itemize}
     \item We propose UnifySpeech, a unified framework for TTS and VC. VC enables unlabeled data to join training process, making TTS encounters more speakers. TTS enhance the ability for voice content decoupling in VC. Thus, both pipeline benefits from the other one.

    \item We apply vector quantization and domain constrain to bridge the gap between the content domains of TTS and VC. Ablation experiment shows this method's effectiveness.

    \item We perform extensive experiments: zero-shot TTS, zero-shot VC. Results proves that jointly trained TTS outperformes StyleSpeech~\citep{min2021meta}and jointly trained VC gains better speech decoupling ability.
\end{itemize}
Demos for this paper are available at \href{https://liuhaogeng.github.io/UnifySpeech/}{https://liuhaogeng.github.io/UnifySpeech/}.

\section{Background}
In this section, we will briefly review the background of this work, including neural TTS and VC models, and the zero-shot learning for TTS and VC tasks.

\subsection{Text-to-speech task}
TTS task is to model the mapping between text and speech, which is a modeling problem between two unequal length sequences. According to the alignment mechanism~\citep{battenberg2020location} between text and speech in the model, the end-to-end TTS model can be divided into two categories: 1) Using a neural network to learn the alignment information between text and speech, such as local sensitive attention in Tacotron~\citep{wang2017tacotron}. Various improvements to the attention mechanism have been proposed. In addition, inspired by CTC~\citep{kim2017joint} in ASR, glow-TTS~\citep{kim2020glow}, VITS~\citep{kim2021conditional} can automatically learn the alignment information. 2) By introducing the duration information~\citep{ren2019fastspeech} of phonemes as prior knowledge, the text is upsampled to achieve alignment~\citep{mcauliffe2017montreal}. Since the upsampled information based on text is independent of the speaker, the speech content can be well separated from the speech. Therefore, we introduce the duration information to build the TTS model in this paper.

\subsection{Voice conversion task}
Voice conversion can be seen as two steps~\citep{sisman2020overview}. Firstly, extract the speaker-independent speech content information from the source speech, and then embed the target speaker information to the speech content to reconstruct the speech of the target speaker. According to the way of extracting speech content, the VC model can be divided into two categories: (1) text-based approach. (2) Information bottleneck approach. The first approach requires an additional pre-trained ASR model. Since the ASR is trained in a supervised manner, it demands a lot of paired text and speech. Additionally, pipeline modeling is easy to accumulate errors and affects the performance of the system. Therefore, a lot of research work is focused on the latter. By adding some restrictions to the information bottleneck, different kinds of information can be decoupled. However, if the information bottleneck can not be decoupled well, the performance of the model will be significantly reduced.

\subsection{Zero shot learning for TTS and VC}
The research of zero-shot learning based on TTS and VC focused on how to extract effective speaker information and then embed it into TTS or VC model for joint training or segmented training. Typical speaker features include i-vector~\citep{wang2017does}, d-vector~\citep{variani2014deep}, x-vector~\citep{snyder2018x} and so on. In addition, the modules that extract speaker-style information through the specially designed network structure, such as GST~\citep{wang2018style}, VAE~\citep{van2017neural}, can also achieve good results.

\section{UnifySpeech}
In this section, we introduce the details of UnifySpeech for zero-shot TTS and VC tasks. We first give the key idea of UnifySpeech: speech factorization, and then introduce the formulation of UnifySpeech. Finally, we will describe the model structure of UnifySpeech.
\begin{figure*}[htp]
    \centering % 图片居中
    \includegraphics[height=4.5cm,width=14cm]{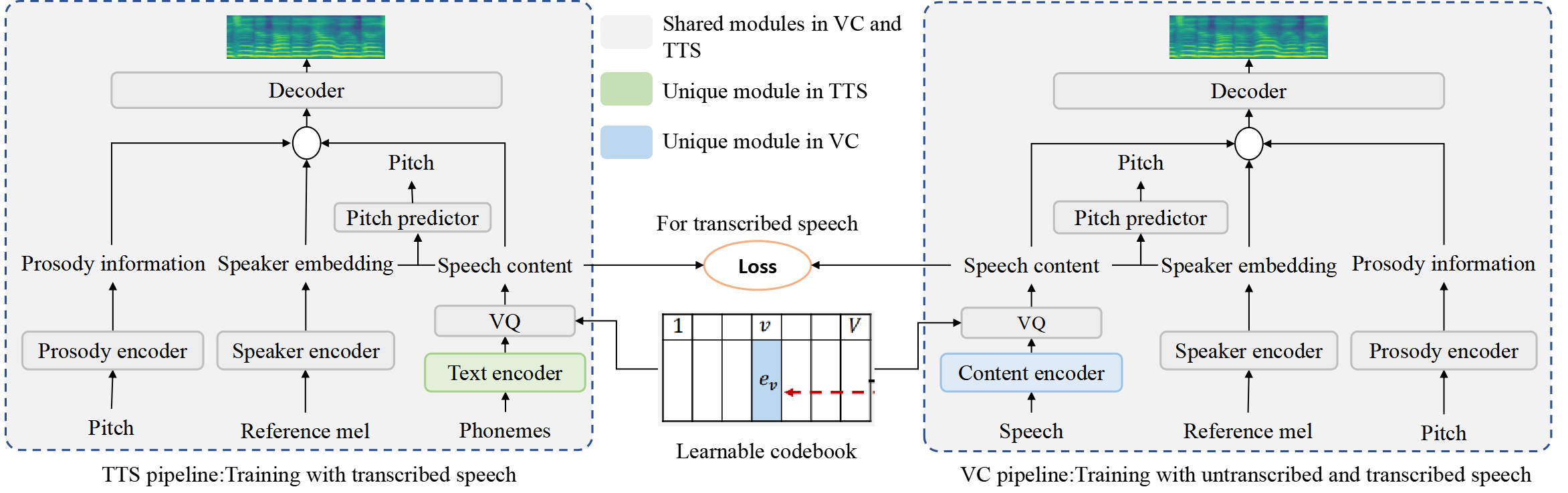}
    \setlength{\abovecaptionskip}{0.3cm}
    \caption{Structure of UnifySpeech.}
    \label{fig:figure1label3}
\vspace{0.3cm}
\end{figure*}

\subsection{Speech representation disentanglement}

The core of the controllable and migratable speech generation task is to decouple the components of the generated speech first, and then control and transfer each component. Although some end-to-end models can directly model the relationship between text and speech (TTS) or speech-to-speech task (VC), due to the mutual coupling of various components of the end-to-end model, it brings great difficulties to the transfer learning of the model. Therefore, we first decouple the speech generation task into three independent components and then input them into the decoder to synthesize speech. This idea is also the main architecture of UnifySpeech. The three components and their sources will be described in detail below.

\textbf{Speech Content}: To generate intelligible speech signals, it is important to model accurate speech content information. Speech content is linguistic information, which is irrelevant to the speaker. Due to the different types of input signals of TTS and VC, the ways of extracting speech content are different. For TTS, the source of speech content is text. Firstly, to learn the context information of the text, a text encoder is used to encode the text to obtain the context representation. The context representation is up-sampled according to the phoneme's duration information to obtain the speech content information. For VC, since the source speech is aligned with the target speech, we directly use a content encoder to extract speech content information from the source speech.

\textbf{Speaker}: The speaker information includes the speaker's characteristics, such as the timbre,  volume, etc. We can extract the speaker information from the given speech of the target speaker. This is common for TTS and VC tasks, so the speaker extraction network can be shared in the two tasks. Since the speaker extraction network can directly extract information from speech without text, so a large number of data without text annotation in the VC task can be used for training, which can help to improve the transfer learning ability of the model.

\textbf{Prosody}: The prosody information represents how the speaker says the content information. It is independent of the speaker information and related to the way of expression. Since the pitch information can reflect the rhythm of speech, therefore, the pitch information is used to extract prosody information. The prosody information, like speaker information, can be shared by both TTS and VC. In addition, in the training process, we can obtain pitch information from the ground truth speech, but there is no ground truth in the inference stage. Therefore, a pitch prediction module~\citep{ren2020fastspeech} is proposed in the training stage, which takes the speech content information and speaker information as the input to predict the pitch information.

\subsection{Speech content with vector quantization}
\label{r2}
Since there are different ways to obtain the speech content in TTS and VC, it is very easy to deviate between the two domains. If this deviation occurs, it will cause devastating damage to some downstream shared modules (such as pitch predictor and decoder). Therefore, to ensure that the consistency of the two speech content domains as close as possible, we reconstruct them from two aspects:
\begin{itemize}
     \item First, we use the shared codebook to quantify the continuous feature space.
        \item Second, we use the labeled data to narrow two discrete feature spaces.
\end{itemize}

The detailed process is described below. Suppose that the vector obtained by the text encoder in TTS pipeline is $C_p = (C_p ^1, C_p ^2, \cdots, C_p ^{T})$ with length $T$, the  vector obtained by the content encoder  in VC pipeline is $C_s = (C_s ^1, C_s ^2, \cdots , C_s ^{T^{'}})$ with length $T^{'}$. It should be noted that we add a length regulator module in the text encoder to solve the problem of length mismatch between the text and speech sequence, which is introduced in FastSpeech~\citep{ren2019fastspeech}. Therefore, if text and speech are paired, $T^{'} = T$. The vector $C_p$ and $C_s$ is a sequence of continuous vector in Eq. Due to the large representation range of continuous features, $C_p$ and $C_s$ are difficult to match. We borrow the discretization method for latent variables from Vector Quantized Variational AutoEncoder~\citep{van2017neural}. Specifically, for each time step $t$, the continuous latent representations $C_p^{t}$ in $C_p$ can be mapped into $\overline{C}_p^{t}$ by finding the nearest pre-defined discretized embedding in the dictionary as:
\begin{align}
\overline{C}_p^{t} = e_k, \quad  k = argmin_{j}  \left\|C_p^{t} - e_j\right\|_2
\end{align}
where $e_j$ is the j-th embedding in the codebook dictionary, and $j \in 1,2, \cdots, V$.  Since selecting the entry with the minimum distance will cause the operation to be non-differentiable, the straight-through gradient estimator can be used to approximate the gradient, which can be expressed as:
\begin{align}
\bar{h}_{t}=h_{t}+e_{v}-\operatorname{sg}\left(h_{t}\right), \quad  v=\underset{k}{\arg \min }\left\|h_{t}-e_{k}\right\|_{2}
\end{align}
where sg(·) is the stop-gradient~\citep{van2017neural} operation that treats its input as constant during back-propagation.

After vector quantization, the quantized sequence  $\overline{C_p} = (\overline{C}_p ^1, \overline{C}_p ^2, \cdots, \overline{C}_p ^{T})$ and $\overline{C_s} = (\overline{C}_s ^1, \overline{C}_s ^2, \cdots, \overline{C}_s^{T})$ can be obtained. It should be noted that when $C_p$ and $C_s$ are quantized into $\overline{C_p}$ and $\overline{C_s}$, they share the same codebook $e$. The advantage of this is that since the speech content features $\overline{C_p}$ in the TTS pipeline are independent of the speaker, sharing the same codebook can help learn the speech content features $\overline{C_s}$ independent of the speaker in the VC pipeline, which is essential for the VC task.

Although both pipeline use the same codebook for coding, there is no guarantee that there is no deviation between the two fields after coding. Therefore, to further eliminate the deviation between the two domains, we use the labeled speech in the TTS pipeline to supervise the training of the two domains. Specifically, for paired text and speech data, we constrain the feature distance between the quantized sequence $\overline{C_p}$ and $\overline{C_s}$:

\begin{align}
\mathcal{L}_{\text {pair}}=\left\|\overline{C_p} -\overline{C_s}\right\|_{2}^{2}
\end{align}
In this way, we can efficiently minimize the domain discrepancy by using limited labeled data.\newline
\begin{figure}[ht]
    \centering % 图片居中
    \includegraphics[width=7cm]{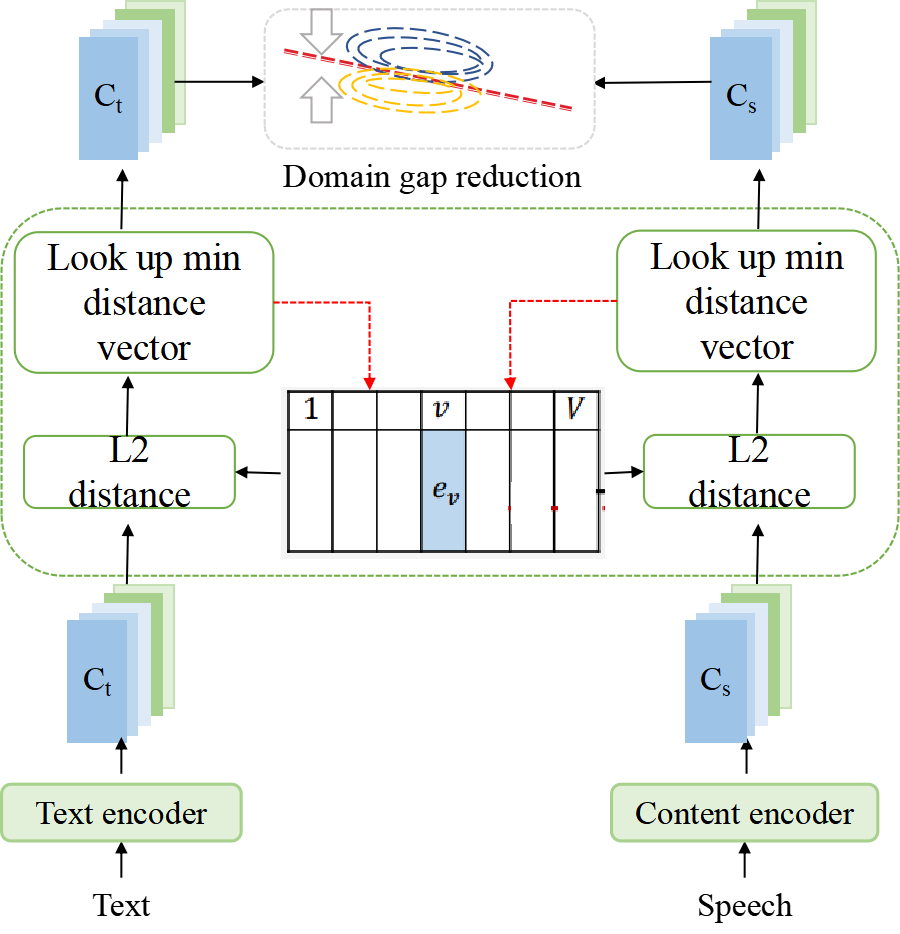}
    \setlength{\abovecaptionskip}{0.3cm}
    \caption{Structure of vector quantized operation.}
    \label{fig:f3}
\vspace{0cm}
\end{figure}

\subsection{UnifySpeech pipeline}
An overview of our proposed UnifySpeech architecture is illustrated in Fig. \ref{fig:f3}. It consists of a sequence-to-sequence TTS, and a sequence-to-sequence VC. The key idea is to share most of the module parameters (speaker encoder, prosody encoder, decoder and pitch predictor) and map the speech content in TTS and VC to the same space. As mentioned above, the UnifySpeech allows us to train the model on the concatenation of both the labeled and unlabeled data. For supervised training with labeled data, both models can be trained independently by minimizing the loss between their predicted speech and the ground truth. For unsupervised training with unlabeled data, the VC pipeline can be trained, and the parameters are shared with TTS.

To further clarify the training process, we unrolled the framework as follows.

\subsubsection{TTS pipeline}

Denote the text and speech sequence pair $(x,y,F0) \in D$, where $D$ is the paired text and speech corpus. Each element in the text sequence $x$ represents a phoneme or character, while each element in the speech sequence $y$ represents a frame of speech. $F0$ is the pitch information of y. The representation obtained after three encoders are speech content $C$, speaker $S$ and prosody $P$.

%\begin{align}
%C &= \text{text\_encoder} ( x ) \\
%S &= \text{speaker\_encoder} ( y ) \\
%p &= \text{prosody\_encoder} ( F0 )
%\end{align}

Then, the three parts are added and input into a decoder to obtain the predicted speech $y'$. In addition, to obtain the pitch information in the interference stage, we use the content information and speaker information to predict the F0. These processes can be expressed as:
\begin{align}
y' &= \text{decoder} (C+S+P) \\
F0' &= \text{pitch\_predictor} ( C+S )
\end{align}
Therefore, the reconstruction loss in TTS process includes two parts:
\begin{align}
\mathcal{L}^{VC}_{rec} = MSE(y,y')  + MSE(F0,F0')
\end{align}

In addition, we use the content encoder in the VC pipeline to extract the content representation $C_s$ for the training speech in TTS, and close the distance between the two domains by calculating the distance loss of $C_s$ and $C_p$, which is explained in Sec. \ref{r2}.
\begin{align}
\mathcal{L}_{\text {pair}}=\left\|\overline{C_p} -\overline{C_s}\right\|_{2}^{2}
\end{align}
The loss of TTS pipeline can be expressed as:
\begin{align}
\mathcal{L}^{TTS} = \mathcal{L}^{TTS}_{rec} + \mathcal{L}_{\text {pair}}
\end{align}
where $MSE$ denotes the mean squared errors.

\subsubsection{VC pipeline}
Denote all the unlabeled or labeled speech $(y,F0) \in Y$. We first extract three information from  $(y,F0)$, which are speech content $C_s$, speaker $S_s$ and prosody $P_s$.

Then, similar to the TTS pipeline, the shared decoder and pitch predictor module is used to predict the speech signal $y'$ and $F0'$, which is similar to the Eq.1 and Eq.2.

The loss of VC pipeline only includes reconstruction loss, which can be expressed as:
\begin{align}
\mathcal{L}^{VC} = MSE(Y,Y') + MSE(F0,F0')
\end{align}
where $MSE$ denotes the mean squared errors.

%\subsubsection{Unified speech content space}
%unlabeled data. For supervised training with label data, bot models can be trained independently by minimizing the loss between their predicted speech and the ground truth. For unsupervised training with unlabeled data, the VC pipeline can be trained, and the parameters are shared with TTS.

%To further clarify the training process, we unrolled the framework as follows.

\subsubsection{Training process}
With such a unified framework, TTS and VC can learn from each other through joint training. The details of the algorithm can be found below.
\begin{algorithm}\label{alg}
	%\textsl{}\setstretch{1.8}
	\renewcommand{\algorithmicrequire}{\textbf{Input:}}
	\renewcommand{\algorithmicensure}{\textbf{Output:}}
	\caption{UnifySpeech training algorithm}
	\label{alg1}
	\begin{algorithmic}[1]
	    \STATE \textbf{Input}: Paired speech and text dataset $(x,y)$, speech only dataset $y^{'}$
	    \REPEAT
		\STATE  \# A. TTS pipeline with speech-text data pairs
		\begin{enumerate}
		\item Sample paired speech and text $(x,y)$
		\item Extract speech content information from $x$ for domain loss
		\item Generate the predict speech y, pitch F0 and speech content from text
		\item Calculate the loss for TTS $\mathcal{L}^{TTS}_{rec}$
		\end{enumerate}
	    \STATE  \# B. VC pipeline with speech-only data
	    \begin{enumerate}
		\item Sample paired speech and text $(x,y)$
		\item Extract speech content information from $y$ for domain loss
		\item Calculate the domain loss for the two pipeline $\mathcal{L}_{pair}$
		\item Sample speech $y^{'}$ in speech only dataset
		\item Generate the predict speech y, pitch F0 and speech content from speech $y^{'}$
		\item Calculate the loss for VC $\mathcal{L}^{VC}$
		\end{enumerate}
		\STATE  \# C. Loss combination:
		\begin{enumerate}
		\item Combine all loss ($\mathcal{L}^{TTS}_{rec}$, $\mathcal{L}_{pair}$, $\mathcal{L}^{VC}$) into a single loss variable
		\item  Calculate TTS and VC parameters gradient
		\item  Update TTS and VC parameters with gradient descent optimization
		\end{enumerate}
		\UNTIL convergence
	\end{algorithmic}
\end{algorithm}

\section{Experiments and results analysis}
In this section, we conduct experiments to evaluate our proposed methods. The experiments are carried out from two aspects: zero-shot TTS, zero-shot VC.

\subsection{Datasets}
  Two datasets are used to simulate labeled data and unlabeled data, respectively. VCTK dataset, an English language dataset containing 44 hours of speech and 109 speakers is used as labeled data. Each speaker has approximately 400 sentences. LibriTTS~\citep{zen2019libritts} are used as unlabeled data, which consists of 585 hours of speech data from 2484 speakers. We only use the speech data in LibriTTS and discard the text for unsupervised training. In this way, it can simulate the scene where a large number of speech that we can obtain are unlabeled. We use a 16-bit, 22050 Hz sampling rate for all experiments. The 80-dim Mel spectrogram is extracted with Hann windowing, frame shift of 12.5-ms, frame length of 50-ms, and 1024-point Fourier transform. In this experiment, we use hifigan~\citep{kong2020hifi} as vocoder.

\subsection{Model Details}
\begin{figure}[H]

    \centering % 图片居中
    \includegraphics[width=10cm]{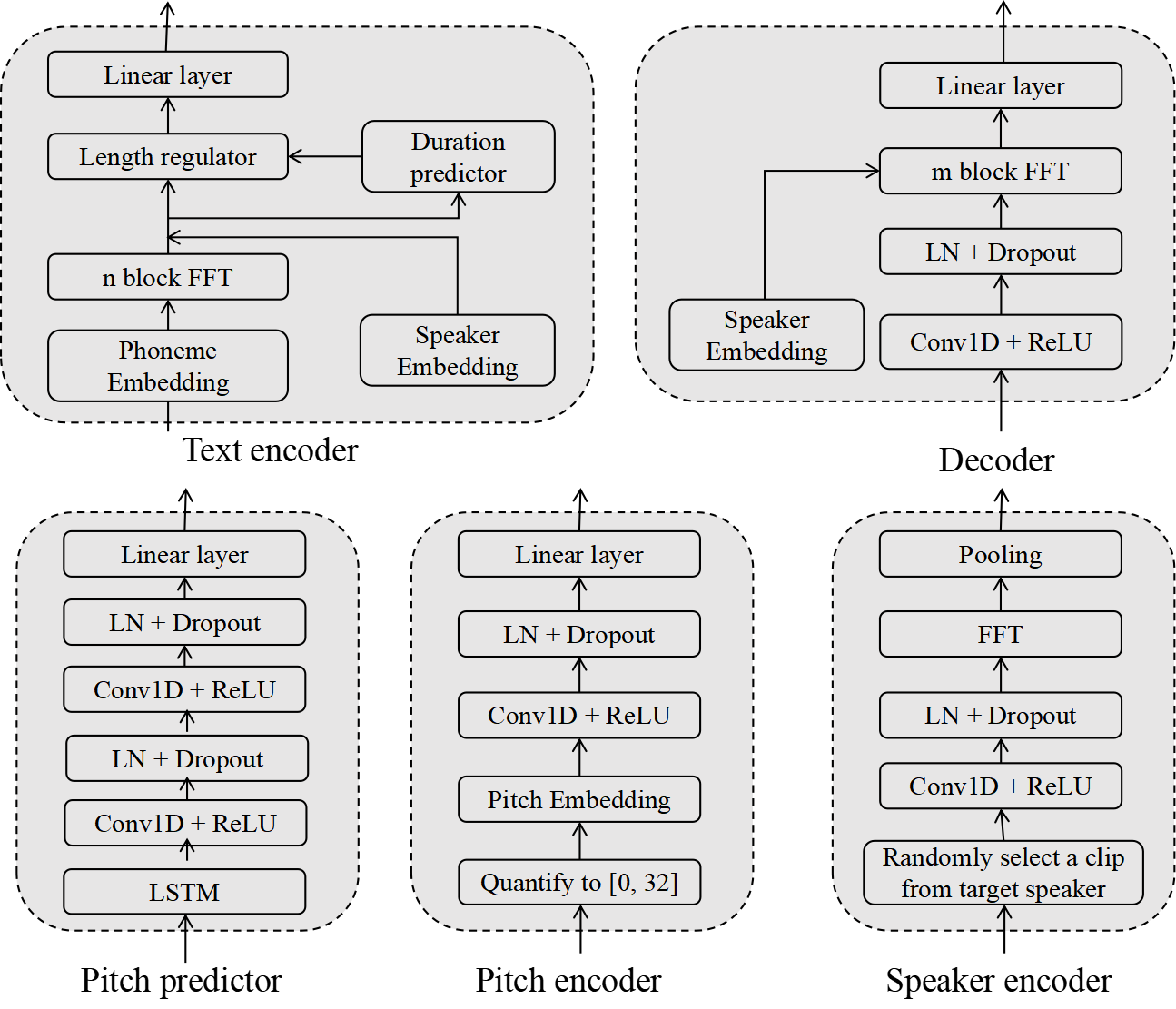}
    \setlength{\abovecaptionskip}{0.3cm}
    \caption{Structure of each module in UnifySpeech. FFT~\citep{ren2019fastspeech} means feed-forward Transformer.}
    \label{fig:detailed}
\end{figure}
The detail of each module in our proposed method is shown in the Fig. \ref{fig:detailed}. Specifically, to make the sequence of speech content extracted from the text encoder and content encoder equal, a length regulator is added to the text encoder, which is inspired by the FastSpeech~\citep{ren2019fastspeech}. The structure of the duration predictor is the same as that in FastSpeech. The structure of the decoder and content encoder is similar, but the dimensions of input and output are opposite. For the prosody encoder, we first quantize F0 of each frame to 32 possible values and encode them into a learnable embedding vector according to the value. And we change the output of the pitch predictor into a distribution. By this way, pitch prediction becomes a classification task, reducing the difficulty of frame-level pitch prediction. We use the speaker encoder in StyleSpeech and fuse features with Style Adaptive Layer Norm(SALN)~\citep{min2021meta} method to make the comparision fairly.

The number of feed-forward Transformer~\citep{vaswani2017attention} (FFT) blocks in the text encoder is 4 and it is 6 in the decoder module. In each FFT block, the dimension of hidden states is 256. The kernel sizes of all the 1D-convolution are set to 3. The dropout rate is set to 0.5. The dimension of the last linear layer in the decoder is 256. The dimension of last linear layer in encoders (text encoder, pitch encoder, content encoder) is 256. An Adam optimizer~\citep{kingma2014adam} is used to update the parameters. The initial learning rate is 0.001 and the learning rate decreased exponentially.
%In this subsection, we will introduce the structure of each module in detail.
\subsection{Zero-shot TTS}
We first carried out zero-shot TTS task. We choose four speakers from VCTK that are not used during training process as target speakers. For each speaker, we randomly select about 20 sentences to be our target. Then we caculate F0 RMSE (root of mean square errors of F0), MCD~\citep{kubichek1993mel} (Mel-cepstrum distortion), V/UV (the error rate of voicing/unvoicing flags) and F0 CORR (correlation factor of F0) between synthesized speech and ground truth speech as the objective metrics.
\begin{table}[H]
\centering
\setlength{\abovecaptionskip}{0.3cm}
\caption{Objective evaluation results for zero-shot TTS.}
\begin{tabular}{ccccc} \hline
\multicolumn{1}{c}{\textbf{Model}}         & \textbf{F0 RMSE (Hz)} & \textbf{MCD (db)} & \multicolumn{1}{c}{\textbf{V/UV}} & \textbf{F0 CORR}  \\
\hline
UnifySpeech                       & \boldsymbol{$17.84$}               & \boldsymbol{$2.51$}            & \boldsymbol{$16.9\%$}                               & \boldsymbol{$0.93$}              \\
\multicolumn{1}{c}{StyleSpeech} & {$19.02$}               & {$2.63$}            & {$18.06\%$}                                & {$0.92$}              \\
\hline
\end{tabular}
\end{table}
Subjective evaluation was also conducted to compare the speech's quality and similarity. We choose mean opinion scores(MOS) for naturalness and similarity mean opinion scores(SMOS) for similarity. Both metrics are rated in 1-to-5 scale and reported
with the 95\% confidence intervals (CI).
\begin{table}[H]

\centering
\setlength{\abovecaptionskip}{0.3cm}
\caption{Mean opinion score (MOS) of the models. With VC means the model is jointly trained.}
\begin{tabular}{c|c|c}
\hline
\textbf{Model}          & \textbf{MOS} & \textbf{SMOS} \\
\hline
GT    & {$4.32\pm0.15$} & {$-$}   \\

GT mel + Vocoder & {$4.09\pm0.15$} & {$-$}  \\
\hline
StyleSpeech    & {$3.52\pm0.13$} & {$3.82\pm0.13$}    \\

UnifySpeech-TTS (with VC) & \boldsymbol{$3.76\pm0.12$}& \boldsymbol{$3.95\pm0.13$}    \\
\hline
\end{tabular}
\end{table}

\begin{figure}[H]
    \centering % 图片居中
    \includegraphics[width=8cm]{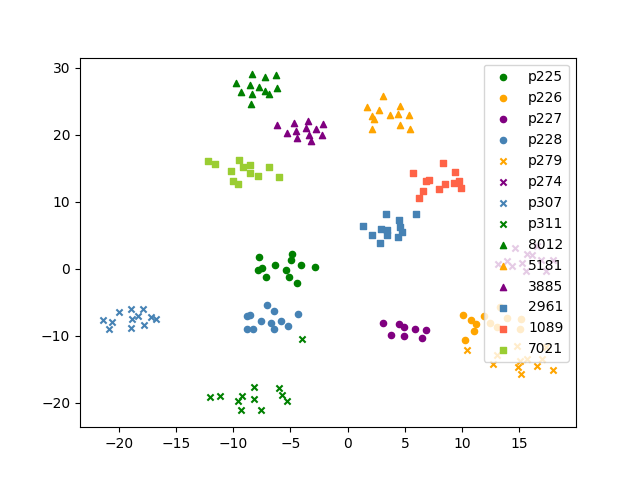}
    \caption{Speaker visualization of generated speeches,where the circle and triangle indicate unseen speaker,while x and square indicate seen speaker.}
    \label{fig:f4}
\end{figure}

To better show our method's effectiveness, we demonstrate the t-SNE projection of the speaker embedding vectors from speakers in both VCTK and LibriTTS.~\citep{van2008visualizing}
Fig. \ref{fig:f4} shows the speaker visualization. For the seen speakers (x) and unseen speakers (o), the corresponding speaker embedding form a cluster and distinct from others. The boundary between different speakers is clear. This shows that the speaker encoder performs well.

\subsection{Zero-shot VC}
We carried out zero-shot VC task, using unseen speakers voice to be the reference speech. As we are lack of parallel corpus, we only conduct subjective evaluation. But VC and TTS shares the same speaker encoder, Fig. \ref{fig:f4} in zero-shot TTS can also be a reference.\newline
Just as in zero-shot TTS task, we choose mean opinion scores(MOS) for naturalness and similarity mean opinion scores(SMOS) for similarity. Both metrics are rated in 1-to-5 scale and reported with the 95\% confidence intervals (CI).

\begin{table}[H]
\centering
\setlength{\abovecaptionskip}{0.3cm}
\caption{Mean opinion score (MOS) of the VC models. With TTS means the model is jointly trained.}
\begin{tabular}{c|c|c}
\hline
\textbf{Model}          & \textbf{MOS} & \textbf{SMOS} \\
\hline
GT    & {$4.32\pm0.15$} & {$-$}   \\

GT mel + Vocoder & {$4.09\pm0.15$} & {$-$}  \\
\hline
UnifySpeech-VC(without TTS) & \boldsymbol{$3.63\pm0.13$}& {$1.31\pm0.06$}    \\
UnifySpeech-VC(with TTS) & {$3.58\pm0.12$}& \boldsymbol{$3.31\pm0.13$}    \\
\hline
\end{tabular}
\end{table}

It can be found that when VC pipeline is trained alone, its performance is poor. In other words, it doesn't have the ability to discriminate. But jointly training with TTS improves its speech decoupling ability, indirectly improving the speaker modeling ability, which we will analysis later.

\subsection{Ablation experiment}
To figure out whether jointly training is effective, we separately train the TTS and VC parts, carrying out objective evaluation on them. We also remove the VQ parts and test the model.
For TTS, we caculate F0 RMSE (root of mean square errors of F0), MCD~\citep{kubichek1993mel} (Mel-cepstrum distortion), V/UV (the error rate of voicing/unvoicing flags) and F0 CORR (correlation factor of F0) between synthesized speech and ground truth speech as the objective metrics.
\begin{table}[H]
\centering
\setlength{\abovecaptionskip}{0.3cm}
\caption{Objective evaluation results for zero-shot TTS.}
\begin{tabular}{ccccc} \hline
\multicolumn{1}{c}{\textbf{Model}}         & \textbf{F0 RMSE (Hz)} & \textbf{MCD (db)} & \multicolumn{1}{c}{\textbf{V/UV}} & \textbf{F0 CORR}  \\
\hline
UnifySpeech-TTS(without VC)                       & {19.31}               & {2.55}            & {16.9\%}                               & {0.91}              \\

UnifySpeech-TTS-novq(with VC) & {$19.41$}                & {$2.58$}            & {$17.2\%$}                                 & {$0.92$}               \\
UnifySpeech-TTS(with VC) & \boldsymbol{$17.84$}                & \boldsymbol{$2.51$}            & \boldsymbol{$16.9\%$}                                 & \boldsymbol{$0.93$}               \\
\hline
\end{tabular}
\end{table}

For VC, we separately calculate the Average Cosine Similarity (ACS)~\citep{lei2022glow} for the embedding from same (S-ACS) and different speakers (D-ACS). And then their ratio is used as an evaluation metric.
\begin{table}[H]
\centering
\setlength{\abovecaptionskip}{0.3cm}
\setlength{\textfloatsep}{0.3cm}

\caption{Ratio of ACS from same and different speaker. Unseen or seen means whether the speaker is from test set. With or without TTS means whether the VC pipeline is trained with TTS pipeline.}
\label{tb:tb2}
\begin{tabular}{|c|c|c|}
\hline
\textbf{Model}            & \textbf{\bm{$\frac{S-ACS}{D-ACS}$}(unseen)} & \textbf{\bm{$\frac{S-ACS}{D-ACS}$}(seen)}  \\
\hline
UnifySpeech-VC (with TTS)    & \boldsymbol{$2.8$}         & \boldsymbol{$6.0$}        \\
\hline
UnifySpeech-VC (without TTS) & 1.0         & 1.0        \\
\hline
\end{tabular}
\end{table}

To validate whether VQ can bridge the gap between the speech content parts in TTS and VC, we caculate the L2 distance between the phoneme representation from TTS and VC. As in VC there are many frames represent same phoneme, so we choose their clustering center as the corresponding phoneme representation. We randomly select 4 sentences from validation set to carry out the experiment.
\begin{table}[H]
\centering
\setlength{\abovecaptionskip}{0.3cm}
\caption{L2 distance between the phoneme representation in TTS and VC. $S_1$ means $sentence_1$.}
\begin{tabular}{c|c|c|c|c|c} \hline
\textbf{Model}        & \textbf{S1} & \textbf{S2} & \textbf{S3} & \textbf{S4} & \textbf{average}  \\
\hline
UnifySpeech-VC(without VQ)                       & {$0.464$}  &    {$0.476$}            & {$0.474$}                               & {$0.553$}   & {$0.492$}           \\\hline
UnifySpeech-VC(with VQ) & \boldsymbol{$0.152$}               & \boldsymbol{$0.205$}            & \boldsymbol{$0.208$}                                & \boldsymbol{$0.205$}    & \boldsymbol{$0.193$}           \\
\hline
\end{tabular}
\end{table}

\subsection{Result analysis}

Above results show that jointly training actually improves TTS's speaker modeling ability and VC's speech decoupling ability. And VQ enhances the consistency of representation of the same content from phonemes and speeches, ensuring the model's correctly working.

For TTS, sharing modules with VC enables unlabeled data to participate in its training process. Along with the richer speaker style pattern, the speaker style modeling capability has been enhanced.

For VC, when trained alone, it doesn't have the speaker modeling ability. As the self-supervised training process aims at reducing the reconstruction loss of source audio, the reference audio of the target speaker isn't crucial in the process. Only the content encoder and the decoder are enough for the process, that's possibly why the speaker embeddings are all similar, though they are from different speakers. When jointly trained, the text loss plays a role of regularization factor, resulting that the content encoder just extracting the content information from the source speech (speaker information is reserved and others are discarded). This makes the reference speech with rich speaker information become indispensable for the reconstruction process. Thus the model's speaker modeling ability has been improved.

\section{Conclusions}
In this paper, we propose UnifySpeech, a unified framework for TTS and VC. Both task benefits from the other one. Due to training with large amounts of unlabeled data, their few-shot modeling ability makes progress as well as the synthesized speech's quality. In the future, further improving the synthesized speech's quality and making the generated speech's style more similar to target speaker will be our endeavor.

\bibliography{iclr2023_conference}
\bibliographystyle{iclr2023_conference}
\end{document}

%% file: math_commands.tex
\usepackage{amsmath,amsfonts,bm}

\def\eqref#1{equation~\ref{#1}}
\def\1{\bm{1}}

\DeclareMathAlphabet{\mathsfit}{\encodingdefault}{\sfdefault}{m}{sl}
\SetMathAlphabet{\mathsfit}{bold}{\encodingdefault}{\sfdefault}{bx}{n}

%% file: iclr2023_conference.bbl
\begin{thebibliography}{25}
\providecommand{\natexlab}[1]{#1}
\providecommand{\url}[1]{\texttt{#1}}
\expandafter\ifx\csname urlstyle\endcsname\relax
  \providecommand{\doi}[1]{doi: #1}\else
  \providecommand{\doi}{doi: \begingroup \urlstyle{rm}\Url}\fi

\bibitem[Battenberg et~al.(2020)Battenberg, Skerry-Ryan, Mariooryad, Stanton,
  Kao, Shannon, and Bagby]{battenberg2020location}
Eric Battenberg, RJ~Skerry-Ryan, Soroosh Mariooryad, Daisy Stanton, David Kao,
  Matt Shannon, and Tom Bagby.
\newblock Location-relative attention mechanisms for robust long-form speech
  synthesis.
\newblock In \emph{ICASSP 2020-2020 IEEE International Conference on Acoustics,
  Speech and Signal Processing (ICASSP)}, pp.\  6194--6198. IEEE, 2020.

\bibitem[Gorodetskii \& Ozhiganov(2022)Gorodetskii and
  Ozhiganov]{gorodetskii2022zero}
Artem Gorodetskii and Ivan Ozhiganov.
\newblock Zero-shot long-form voice cloning with dynamic convolution attention.
\newblock \emph{arXiv preprint arXiv:2201.10375}, 2022.

\bibitem[Kim et~al.(2020)Kim, Kim, Kong, and Yoon]{kim2020glow}
Jaehyeon Kim, Sungwon Kim, Jungil Kong, and Sungroh Yoon.
\newblock Glow-tts: A generative flow for text-to-speech via monotonic
  alignment search.
\newblock \emph{Advances in Neural Information Processing Systems},
  33:\penalty0 8067--8077, 2020.

\bibitem[Kim et~al.(2021)Kim, Kong, and Son]{kim2021conditional}
Jaehyeon Kim, Jungil Kong, and Juhee Son.
\newblock Conditional variational autoencoder with adversarial learning for
  end-to-end text-to-speech.
\newblock In \emph{International Conference on Machine Learning}, pp.\
  5530--5540. PMLR, 2021.

\bibitem[Kim et~al.(2017)Kim, Hori, and Watanabe]{kim2017joint}
Suyoun Kim, Takaaki Hori, and Shinji Watanabe.
\newblock Joint ctc-attention based end-to-end speech recognition using
  multi-task learning.
\newblock In \emph{2017 IEEE international conference on acoustics, speech and
  signal processing (ICASSP)}, pp.\  4835--4839. IEEE, 2017.

\bibitem[Kingma \& Ba(2014)Kingma and Ba]{kingma2014adam}
Diederik~P Kingma and Jimmy Ba.
\newblock Adam: A method for stochastic optimization.
\newblock \emph{arXiv preprint arXiv:1412.6980}, 2014.

\bibitem[Kong et~al.(2020)Kong, Kim, and Bae]{kong2020hifi}
Jungil Kong, Jaehyeon Kim, and Jaekyoung Bae.
\newblock Hifi-gan: Generative adversarial networks for efficient and high
  fidelity speech synthesis.
\newblock \emph{Advances in Neural Information Processing Systems},
  33:\penalty0 17022--17033, 2020.

\bibitem[Kubichek(1993)]{kubichek1993mel}
Robert Kubichek.
\newblock Mel-cepstral distance measure for objective speech quality
  assessment.
\newblock In \emph{Proceedings of IEEE pacific rim conference on communications
  computers and signal processing}, volume~1, pp.\  125--128. IEEE, 1993.

\bibitem[Lei et~al.(2022)Lei, Yang, Cong, Xie, and Su]{lei2022glow}
Yi~Lei, Shan Yang, Jian Cong, Lei Xie, and Dan Su.
\newblock Glow-wavegan 2: High-quality zero-shot text-to-speech synthesis and
  any-to-any voice conversion.
\newblock \emph{arXiv preprint arXiv:2207.01832}, 2022.

\bibitem[McAuliffe et~al.(2017)McAuliffe, Socolof, Mihuc, Wagner, and
  Sonderegger]{mcauliffe2017montreal}
Michael McAuliffe, Michaela Socolof, Sarah Mihuc, Michael Wagner, and Morgan
  Sonderegger.
\newblock Montreal forced aligner: Trainable text-speech alignment using kaldi.
\newblock In \emph{Interspeech}, volume 2017, pp.\  498--502, 2017.

\bibitem[Min et~al.(2021)Min, Lee, Yang, and Hwang]{min2021meta}
Dongchan Min, Dong~Bok Lee, Eunho Yang, and Sung~Ju Hwang.
\newblock Meta-stylespeech: Multi-speaker adaptive text-to-speech generation.
\newblock In \emph{International Conference on Machine Learning}, pp.\
  7748--7759. PMLR, 2021.

\bibitem[Ren et~al.(2019)Ren, Ruan, Tan, Qin, Zhao, Zhao, and
  Liu]{ren2019fastspeech}
Yi~Ren, Yangjun Ruan, Xu~Tan, Tao Qin, Sheng Zhao, Zhou Zhao, and Tie-Yan Liu.
\newblock Fastspeech: Fast, robust and controllable text to speech.
\newblock \emph{Advances in Neural Information Processing Systems}, 32, 2019.

\bibitem[Ren et~al.(2020)Ren, Hu, Tan, Qin, Zhao, Zhao, and
  Liu]{ren2020fastspeech}
Yi~Ren, Chenxu Hu, Xu~Tan, Tao Qin, Sheng Zhao, Zhou Zhao, and Tie-Yan Liu.
\newblock Fastspeech 2: Fast and high-quality end-to-end text to speech.
\newblock \emph{arXiv preprint arXiv:2006.04558}, 2020.

\bibitem[Sisman et~al.(2020)Sisman, Yamagishi, King, and
  Li]{sisman2020overview}
Berrak Sisman, Junichi Yamagishi, Simon King, and Haizhou Li.
\newblock An overview of voice conversion and its challenges: From statistical
  modeling to deep learning.
\newblock \emph{IEEE/ACM Transactions on Audio, Speech, and Language
  Processing}, 29:\penalty0 132--157, 2020.

\bibitem[Snyder et~al.(2018)Snyder, Garcia-Romero, Sell, Povey, and
  Khudanpur]{snyder2018x}
David Snyder, Daniel Garcia-Romero, Gregory Sell, Daniel Povey, and Sanjeev
  Khudanpur.
\newblock X-vectors: Robust dnn embeddings for speaker recognition.
\newblock In \emph{2018 IEEE international conference on acoustics, speech and
  signal processing (ICASSP)}, pp.\  5329--5333. IEEE, 2018.

\bibitem[Tang et~al.(2021)Tang, Luo, Zhao, Yin, Zhao, and Zeng]{tang2021zero}
Chuanxin Tang, Chong Luo, Zhiyuan Zhao, Dacheng Yin, Yucheng Zhao, and Wenjun
  Zeng.
\newblock Zero-shot text-to-speech for text-based insertion in audio narration.
\newblock \emph{arXiv preprint arXiv:2109.05426}, 2021.

\bibitem[Van Den~Oord et~al.(2017)Van Den~Oord, Vinyals, et~al.]{van2017neural}
Aaron Van Den~Oord, Oriol Vinyals, et~al.
\newblock Neural discrete representation learning.
\newblock \emph{Advances in neural information processing systems}, 30, 2017.

\bibitem[Van~der Maaten \& Hinton(2008)Van~der Maaten and
  Hinton]{van2008visualizing}
Laurens Van~der Maaten and Geoffrey Hinton.
\newblock Visualizing data using t-sne.
\newblock \emph{Journal of machine learning research}, 9\penalty0 (11), 2008.

\bibitem[Variani et~al.(2014)Variani, Lei, McDermott, Moreno, and
  Gonzalez-Dominguez]{variani2014deep}
Ehsan Variani, Xin Lei, Erik McDermott, Ignacio~Lopez Moreno, and Javier
  Gonzalez-Dominguez.
\newblock Deep neural networks for small footprint text-dependent speaker
  verification.
\newblock In \emph{2014 IEEE international conference on acoustics, speech and
  signal processing (ICASSP)}, pp.\  4052--4056. IEEE, 2014.

\bibitem[Vaswani et~al.(2017)Vaswani, Shazeer, Parmar, Uszkoreit, Jones, Gomez,
  Kaiser, and Polosukhin]{vaswani2017attention}
Ashish Vaswani, Noam Shazeer, Niki Parmar, Jakob Uszkoreit, Llion Jones,
  Aidan~N Gomez, {\L}ukasz Kaiser, and Illia Polosukhin.
\newblock Attention is all you need.
\newblock \emph{Advances in neural information processing systems}, 30, 2017.

\bibitem[Wang et~al.(2017{\natexlab{a}})Wang, Qian, and Yu]{wang2017does}
Shuai Wang, Yanmin Qian, and Kai Yu.
\newblock What does the speaker embedding encode?
\newblock In \emph{Interspeech}, pp.\  1497--1501, 2017{\natexlab{a}}.

\bibitem[Wang et~al.(2017{\natexlab{b}})Wang, Skerry-Ryan, Stanton, Wu, Weiss,
  Jaitly, Yang, Xiao, Chen, Bengio, et~al.]{wang2017tacotron}
Yuxuan Wang, RJ~Skerry-Ryan, Daisy Stanton, Yonghui Wu, Ron~J Weiss, Navdeep
  Jaitly, Zongheng Yang, Ying Xiao, Zhifeng Chen, Samy Bengio, et~al.
\newblock Tacotron: Towards end-to-end speech synthesis.
\newblock \emph{arXiv preprint arXiv:1703.10135}, 2017{\natexlab{b}}.

\bibitem[Wang et~al.(2018)Wang, Stanton, Zhang, Ryan, Battenberg, Shor, Xiao,
  Jia, Ren, and Saurous]{wang2018style}
Yuxuan Wang, Daisy Stanton, Yu~Zhang, RJ-Skerry Ryan, Eric Battenberg, Joel
  Shor, Ying Xiao, Ye~Jia, Fei Ren, and Rif~A Saurous.
\newblock Style tokens: Unsupervised style modeling, control and transfer in
  end-to-end speech synthesis.
\newblock In \emph{International Conference on Machine Learning}, pp.\
  5180--5189. PMLR, 2018.

\bibitem[Yan et~al.(2021)Yan, Tan, Li, Qin, Zhao, Shen, and
  Liu]{yan2021adaspeech}
Yuzi Yan, Xu~Tan, Bohan Li, Tao Qin, Sheng Zhao, Yuan Shen, and Tie-Yan Liu.
\newblock Adaspeech 2: Adaptive text to speech with untranscribed data.
\newblock In \emph{ICASSP 2021-2021 IEEE International Conference on Acoustics,
  Speech and Signal Processing (ICASSP)}, pp.\  6613--6617. IEEE, 2021.

\bibitem[Zen et~al.(2019)Zen, Dang, Clark, Zhang, Weiss, Jia, Chen, and
  Wu]{zen2019libritts}
Heiga Zen, Viet Dang, Rob Clark, Yu~Zhang, Ron~J Weiss, Ye~Jia, Zhifeng Chen,
  and Yonghui Wu.
\newblock Libritts: A corpus derived from librispeech for text-to-speech.
\newblock \emph{arXiv preprint arXiv:1904.02882}, 2019.

\end{thebibliography}
